\documentclass{elsart}
\usepackage{amsmath}

\usepackage{graphicx}

\usepackage{amssymb}

\begin{document}

\def\d{\partial}
\def\um{\,\mu{\rm  m}}
\def\mm{\,   {\rm mm}}
\def\cm{\,   {\rm cm}}
\def \m{\,   {\rm  m}}
\def\ps{\,   {\rm ps}}
\def\ns{\,   {\rm ns}}
\def\us{\,\mu{\rm  s}}
\def\ms{\,   {\rm ms}}
\def\nA{\,   {\rm nA}}
\def\uA{\,\mu{\rm  A}}
\def\mA{\,   {\rm mA}}
\def\A {\,   {\rm  A}}
\def\mV{\,   {\rm mV}}
\def\V {\,   {\rm  V}}
\def\fF{\,   {\rm fF}}
\def\pF{\,   {\rm pF}}
\def\GeV{\, {\rm GeV}}
\def\MHz{\, {\rm MHz}}
\def\uW{\,\mu{\rm  W}}
\def\e {\,  {\rm e^-}}

\renewcommand{\labelenumi}{\arabic{enumi}}
\renewcommand{\labelitemi}{-}

\begin{frontmatter}

\title{2D Detectors for Particle Physics and for Imaging Applications\thanksref{BMBF} }

\thanks[BMBF]{Work supported by the German Ministerium f{\"u}r Bildung,
              und Forschung (BMBF) under contract
              no.~$05 HA1PD1/5$\ , by the
              Ministerium f{\"u}r Wissenschaft und Forschung (MWF) des Landes
              Nordrhein--Westfalen under contract no.~$IV\,A5-106\,011\,98$, and
              by the DIP Foundation under contract no. E7.1}

\author {H.~Kr\"uger\thanksref{HK}}

\thanks[HK]{Physikalisches Institut, Nussallee 12,
               D-53115 Bonn, Germany, Tel.: +49\,228\,73-2996, Fax:
               -3220, email: krueger@physik.uni-bonn.de
           }
\address{Physikalisches Institut der Universit{\"a}t Bonn, Germany}

\begin{abstract}
The demands on detectors for particle detection as well as for
medical and astronomical X-ray imaging are continuously pushing the 
development of novel pixel detectors. The state of the art in pixel 
detector technology to date are hybrid pixel detectors in which sensor 
and read-out integrated circuits are processed on different substrates 
and connected via high density interconnect structures. While these 
detectors are technologically mastered such that large scale particle
detectors can be and are being built, the demands for improved performance
for the next generation particle detectors ask for the development of
monolithic or semi-monolithic approaches. Given the fact that the demands
 for medical imaging are different in some key aspects, developments
for these applications, which started as particle physics spin-off, are
becomming rather independent. New approaches are leading to novel signal
processing concepts and interconnect technologies to satisfy the need for
very high dynamic range and large area detectors. The present state in
hybrid and (semi-)monolithic pixel detector development and their different
approaches for particle physics and imaging application is reviewed.
\end{abstract}

\end{frontmatter}
\newpage

\section{Hybrid Pixel Detectors for tracking applications}
The development of \emph{hybrid pixel detectors} over the last 10 
years was mainly pushed by the specifications for the vertex detectors for 
the large high energy physics experiments ALICE \cite{ALICE_pix,ALICE_Riedler}, 
ATLAS \cite{ATLAS_pix,ATLAS_Gemme}, CMS \cite{CMS_pix,CMS_Erdmann},
LHCb \cite{LHCb_pix} and fixed target experiment NA60 \cite{NA60,NA60_Radermacher} 
at the Large Hadron Collider (LHC) at CERN and the BTeV detector at the TEVATRON \cite{BTEV_pix}.
All these experiments have a high demand on spatial resolution, timing
precision and radiation tolerance. Also the feasibility of building 
large detector areas (up to $\sim$2m$^2$) had to be proven. The 
hybrid pixel technology, where electronic chip and sensor elements
are on different substrates which are connected via flip-chip
assembly, showed to be mature enough to comply with all these demands.
The pixel detectors, which are closest to the interaction region of 
a collider experiment, comprise of typically 2 to 3 cylindrical layers 
and additional disk layers for the forward and backward region. The
pixel sizes are between 50$\mu$m $\times$ 400$\mu$m and 
100$\mu$m $\times$ 150$\mu$m for the mentioned experiments to achieve
the spatial resolution which is required for efficient identification
of short lived particles (b-tagging for Higgs and SUSY signals) and 
robust event reconstruction at high luminosities.

\begin{figure}[h]
\begin{center}
\includegraphics[width=0.35\textwidth]{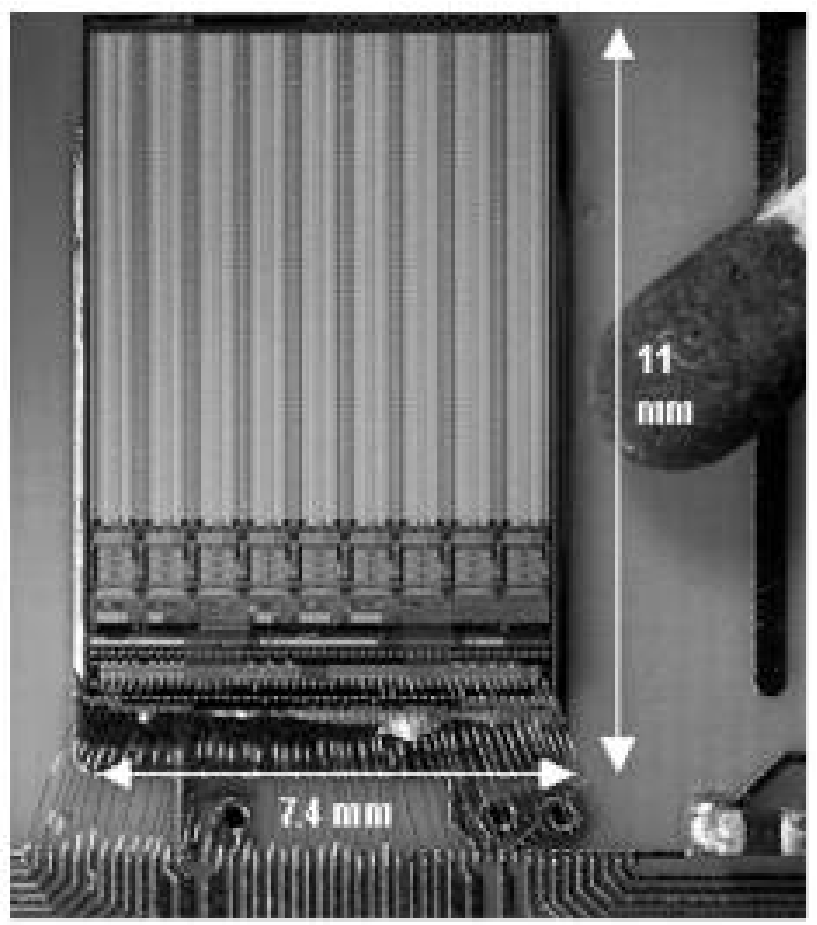}
\includegraphics[width=0.4\textwidth]{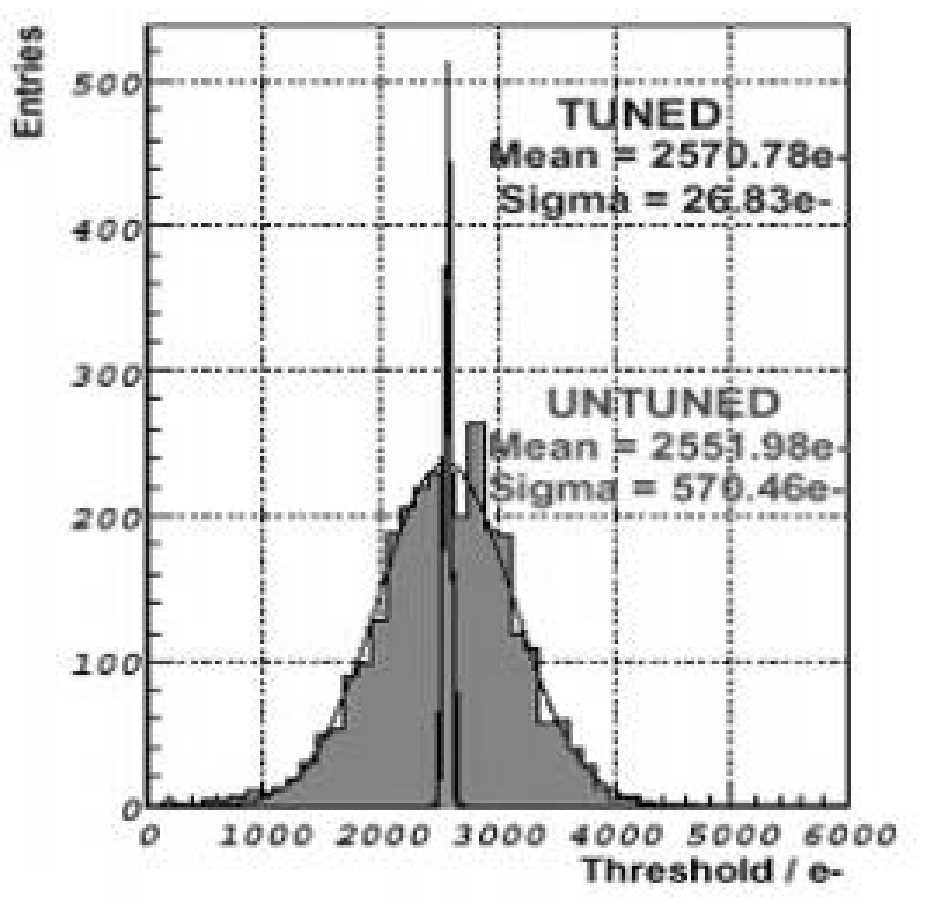}
\end{center}
\caption[]{Prototype of the ATLAS front-end chip (left). 
Dispersion of the pixel thresholds before and after tunning (right).} 
\label{FE_chip}
\end{figure}

Apart from the electronical requirements on the pixel electronics \cite{blanquart03}
like low power consumption ($<$ 50$\mu$W per pixel), low noise and
threshold dispersion (together $<$~200e), zero suppression in
every pixel, on-chip hit buffering and sufficient timing precision
for LHC bunch crossing rates (25 ns), the essential radiation hardness has been 
a major challenge for the design of front-end chips. This could be
achieved with special design techniques and the use of 
deep submicron CMOS technologies, which superseeded dedicated 
radhard processes. Figure \ref{FE_chip} shows a picture of a prototype
ATLAS pixel chip and a measurement of the threshold distribution which
is a measure for the homogeneity of the chip response. The dispersion 
of about 600 e$^-$ can be lowered to below 50 e$^-$ by a 7-bit 
tuning feature implemented in the chip. This is well below the
electronic noise of about 170 e$^-$. However the minimum operational
threshold is limited to $>$~1500 e due to the crosstalk between the digital 
and the analog part of the pixel chip.

\begin{figure}[htb]
\begin{center}
\includegraphics[width=0.6\textwidth]{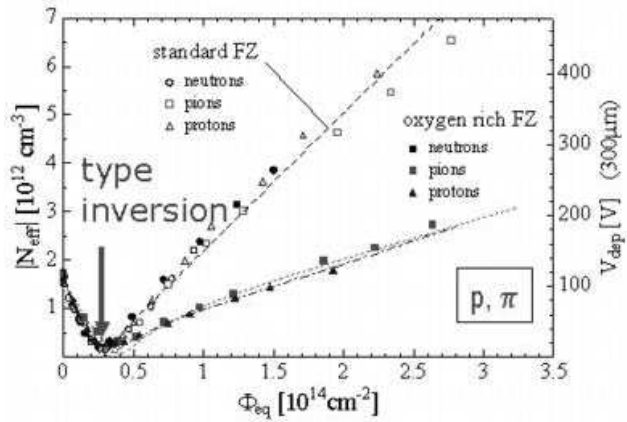}
\end{center}
\caption[]{Depletion voltage and effective dopant concentration $N_{eff}$ versus 1 MeV neutron equivalent flux.} \label{OxSi-Sensor}
\end{figure}

Figure \ref{OxSi-Sensor} shows the improvement of the radiation tolerance 
of the sensor material by the use of oxygenated silicon which has 
less sensitivity to non-ionizing energy loss of charged particles \cite{oxysilicon}. 
The fact that radiation damage caused by neutrons is the same as with standard
silicon is not fully understood yet.
The sensor pixels are implanted as $n^+$ electrodes in n-bulk material.
After type inversion which occurs after about $\Phi_{eq} = 2.5 \times
10^{13}$cm$^{-2}$ the diode junction is at the electrode side, allowing
the partial depleted operation of the detector with less than 600 V after
ten years of LHC operation.

\begin{figure}[htb]
\includegraphics[width=0.355\textwidth]{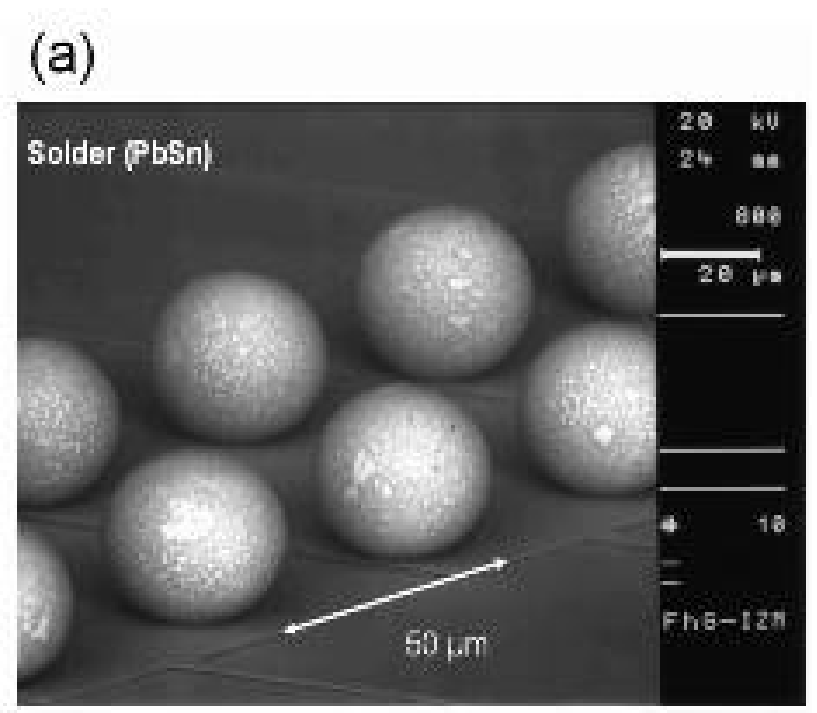}
\includegraphics[width=0.42\textwidth]{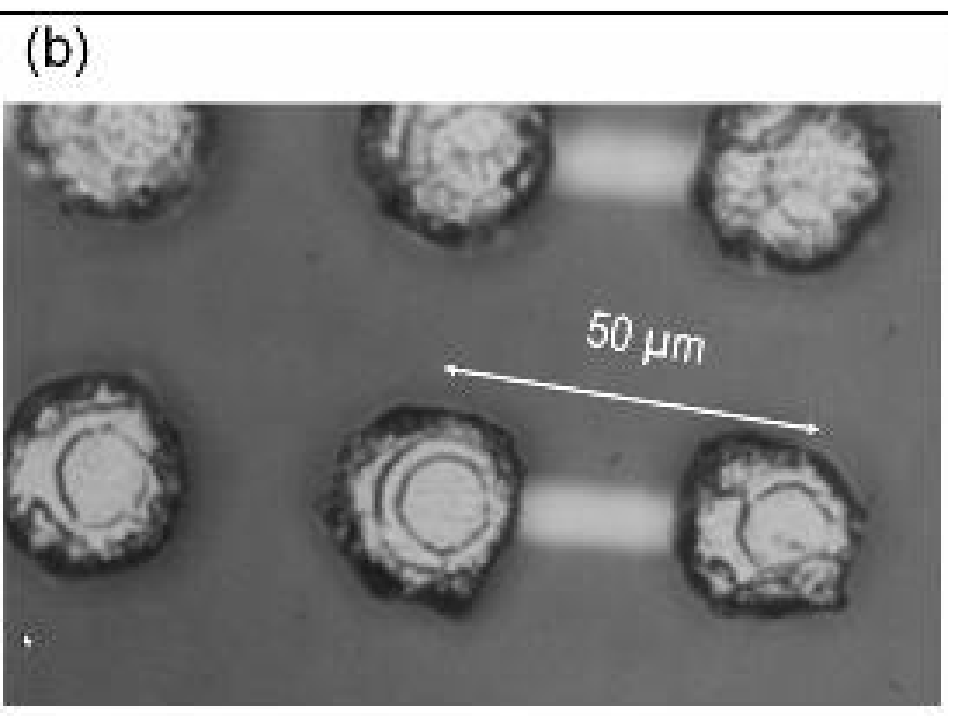}
\includegraphics[width=0.168\textwidth]{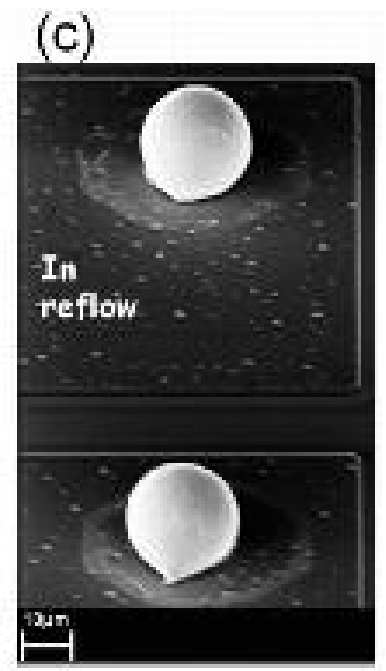}
\caption{(a) solder (PbSn, Photo IZM, Berlin) (b) Indium (Photo
AMS, Rome), and (c) Indium with reflow (Photo PSI, Villingen) bump
rows with $50 \mu$m pitch.} \label{bumps}
\end{figure}

\begin{figure}[htb]
\includegraphics[width=0.35\textwidth]{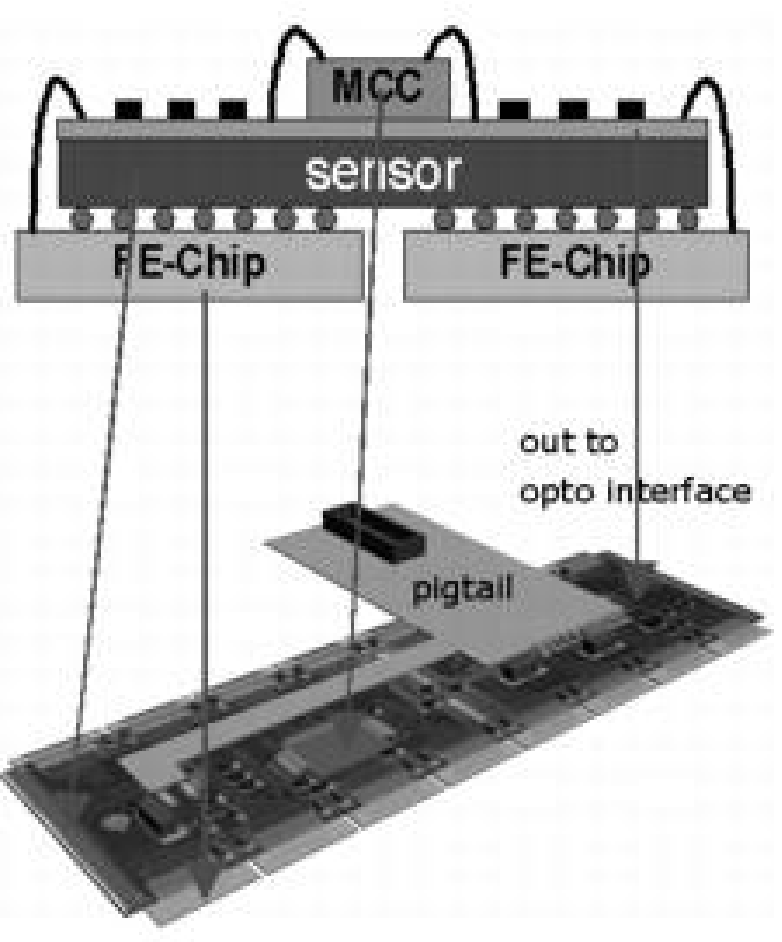}
\includegraphics[width=0.6\textwidth]{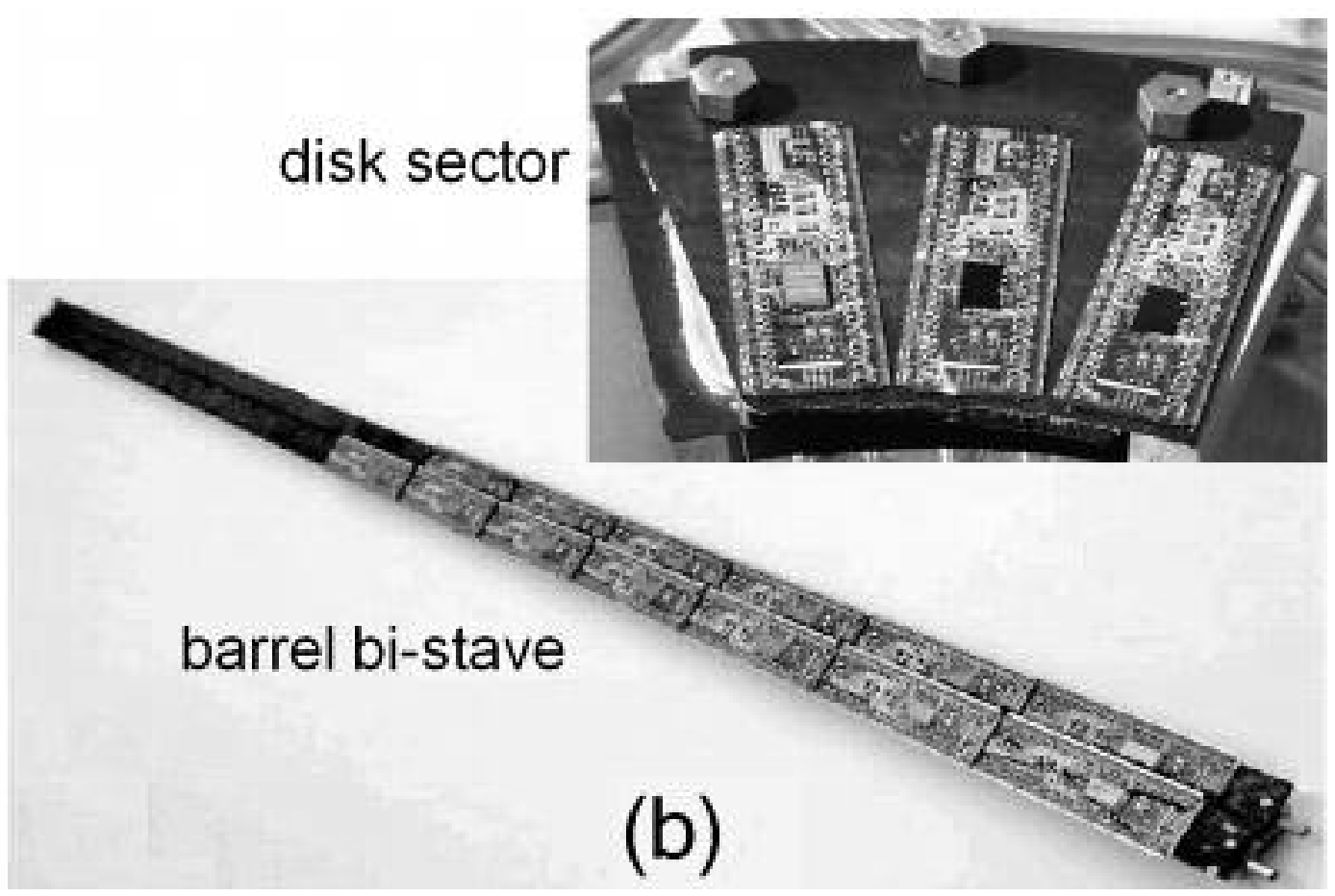}
  \caption{Cross-section of a typical module assembly (left) and
   ATLAS modules mounted to a bi-stave unit and to a disk sector (right).}
  \label{modules}
\end{figure}

The assembly of the sensor \emph{modules} is done by fine pitch 
flip-chip bonding the electronic chips and the sensor. Mainly two different
technologies are used to build sensor modules in large scales: Solder 
(PbSn) bump bonding \cite{PbSn_bumping} with reflow and Indium bump bonding 
\cite{Indium_bumping} with optional reflow \cite{In-reflow_bumping}
and thermal compression. Fig. \ref{bumps} shows rows of $50 \mu$m pitch bumps 
obtained by these techniques. All technologies have been successfully used with 8"
IC-wafers and 4" sensor wafers. A module for the ATLAS or CMS pixel detector is 
composed of 16 front-end chips bump-bonded to one silicon sensor with an area of 
typically 2 cm $\times$ 6.5 cm. As shown in Fig. \ref{modules} a kapton flex
circuit which supports a module control chip is glued on the backside of 
the sensor. Wire bonds connect the front-end chip I/O lines and a high
density interconnect adapter (pigtail) to the flex circuit.

To minimize the radiation induced damage of the sensors due to the harsh 
radiation environment the mechanical support structure for the modules 
also has to provide cooling to temperatures below the freezing point. To 
compensate this additional material budget the front-end chips are thinned 
down to 180 $\mu$m giving a total thickness of the module at normal incidence 
of about $2 - 3 \%$ $X_0$ ($X_0$ is the radiation length which specifies
the average lenght of path in a specific material in which a relativistic 
charged particle will loose 67~$\%$ of its energy). 

\section{Imaging with Hybrid Pixel Detectors}
With the design of pixel chip electronics for vertex trackers
the capability of detecting individual radiation quanta
made hybrid pixel detectors also very attractive for non-HEP 
applications like medical X-ray imagers or synchrotron radiation detectors.
The counting principle leads to superior performance of such detectors 
compared to standard film-foil or scintillator-CCD systems 
that are normally used for imaging: an in principal unlimited dynamic 
range and full linearity in their response function.

While in tracking applications the timing information (and optional the
energy loss) of an incident particle is measured within every pixel,
most of the present imaging applications are based on counting the individual quanta for
a certain exposure time intervall. Therefore the first implementations 
of \emph{counting pixel detectors} simply replaced the digital part of 
the pixel electronic of a vertex tracker pixel chip, which generates the 
timestamp for each individual event, by a counter \cite{MEDIPIX1}\cite{MPEC_ref}.
The same principle is also used for protein-crystallography with synchrotron radiation
\cite{Graafsma_Portland,3Dwestbrook}. But for successful development of an 
optimized imaging system based on a hybrid pixel detector one has to obey 
the different demands for both application areas. Table \ref{table1} gives
a summary of the different system aspects and subsequent demands on the 
pixel electronics and sensor charcteristics for tracking and imaging 
applications, respectively.

\begin{table}[htb]
\includegraphics[width=1\textwidth]{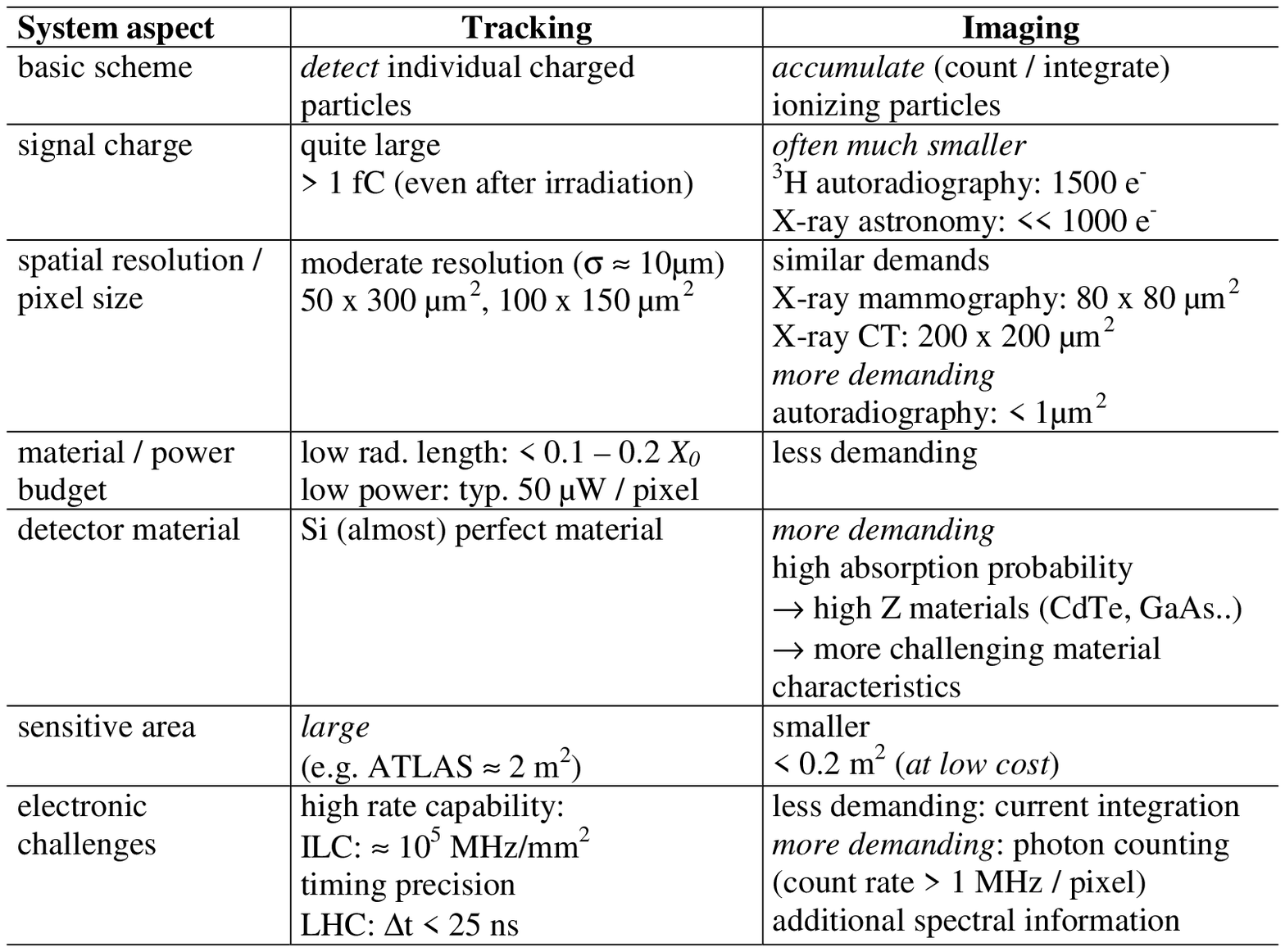}
  \caption{Comparison of the demands on pixel detectors for tracking and imaging applications}
  \label{table1}
\end{table}

Conventional X-ray imagers integrate the energy of the absorbed quanta. The so 
called \emph{Active Matrix Flat Panel Imager} \cite{flatpanel1,flatpanel2,a-Se,flatpanel4}, 
which represents the state-of-the-art in digital X-ray detectors, exploits two different detection
methods: indirect conversion, where a scintillator (e.g. CsI) is deposited atop
a TFT photo diode array, and direct conversion where an absorber (e.g. Selenium) converts the 
incident X-rays to a charge which is collected with an TFT capacitor array \cite{flatpanel3}.
Due to the principle of accumulating the incident particles, the rate capability
of integrating imagers is practically unlimited. To be competitive with these detectors
the counting approach needs to meet the high count rates 
($\sim$ 10 MHz/mm$^2$) and the high dynamic range of at least 15 bit. In addition
these detectors should combine low noise and low thresholds with
a very low threshold dispersion to allow homogenous imaging.

\subsection*{Medical Imaging}
Besides the superior signal to noise figures, which reach the quantum limit,
the counting principle offers the fundamental advantage of attaining and
exploiting an additional spectral information from the incident X-ray photons.
In the simplest implementation a differential energy measurement is realized 
with a double threshold \cite{MPEC-windowing1,MPEC-windowing2,MEDIPIX2}
which can enhance the contrast of an image as the X-ray energy 
spectrum is different behind different absorbers (e.g. bone or soft tissue). 
The next step towards the use of the spectral information could be the implementation
of more discrete energy bins and the use of an energy weighting technique
which has the potential to enhance the image contrast \cite{MEDIPIX_EWT}.
A homogeneous response of the absorber material, low threshold dispersion and control of the 
charge shared between adjacent pixels is crucial however. Another issue -- especially with
the dense electronics of a hybrid pixel detector -- is the digital noise (crosstalk between the digital and the analog part), which is still the limitinig factor for a
low threshold operation of a hybrid pixel detector system.

Also a high photon absorption efficiency 
is mandatory for radiography requiring the use of high-Z sensor materials like CdTe or CZT.
With these materials the development of large area detectors and their hybridization is still
a challenging task.

\begin{figure}
\begin{center}
\includegraphics[width=0.9\textwidth]{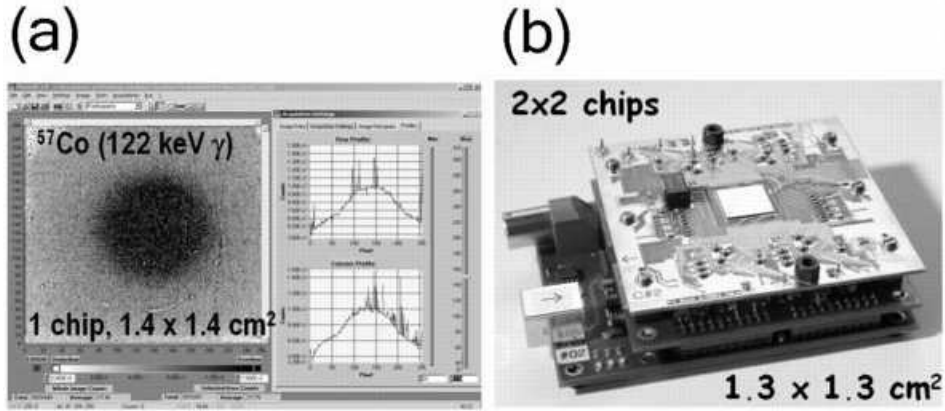}
\end{center}
\caption[]{(a) Image of a $^{57}$Co (122 keV $\gamma$) point
source taken with a MEDIPIX2 counting single chip module (14x14
mm$^2$)\cite{MEDIPIX_Portland}. (b) MPEC 2x2 multi chip module
with a CdTe sensor \cite{MPEC_Portland}.} \label{counting}
\end{figure}

There are serveral counting pixel system development efforts
for medical applications. Among those projects two chip developments
exhibit a rather mature R\&D status: the MEDIPIX collaboration \cite{MEDIPIX,MEDIPIX2}
uses the MEDIPIX2 chip with 256$\times$256, 55$\times$55$\mu$m$^2$ 
pixels fabricated in $0.25 \mu$m technology, energy
windowing via two tunable discriminator thresholds, and a 13 bit
counter. The maximum count rate per pixel is about 1 MHz. Fig.
\ref{counting}(a) shows an image of a $^{57}$Co (122 keV $\gamma$)
1 mm diameter point source obtained with the Medipix2 single chip
bonded to a 14x14 mm$^2$ CdTe sensor \cite{MEDIPIX_Portland}. A
Multi-Chip module with 2x2 chips using high-Z CdTe sensors with
the MPEC chip \cite{MPEC_Portland} is shown in Fig.
\ref{counting}(b). The MPEC chip features $32 \times 32$ pixels
(200$\times$200$\mu$m$^2$), double threshold operation, 18-bit
counting at $\sim$1 MHz per pixel as well as low noise values
($\sim$120e with CdTe sensor) and threshold dispersion ($21$e
after tuning) \cite{MPEC_ref,MPEC_Portland}. A technical issue
here is the bumping of individual die CdTe sensors which has been
solved using Au-stud bumping with In-topping \cite{MPEC_CdTe}.

\subsection*{Protein Crystallography}
\begin{figure}
\begin{center}
\includegraphics[width=0.65\textwidth]{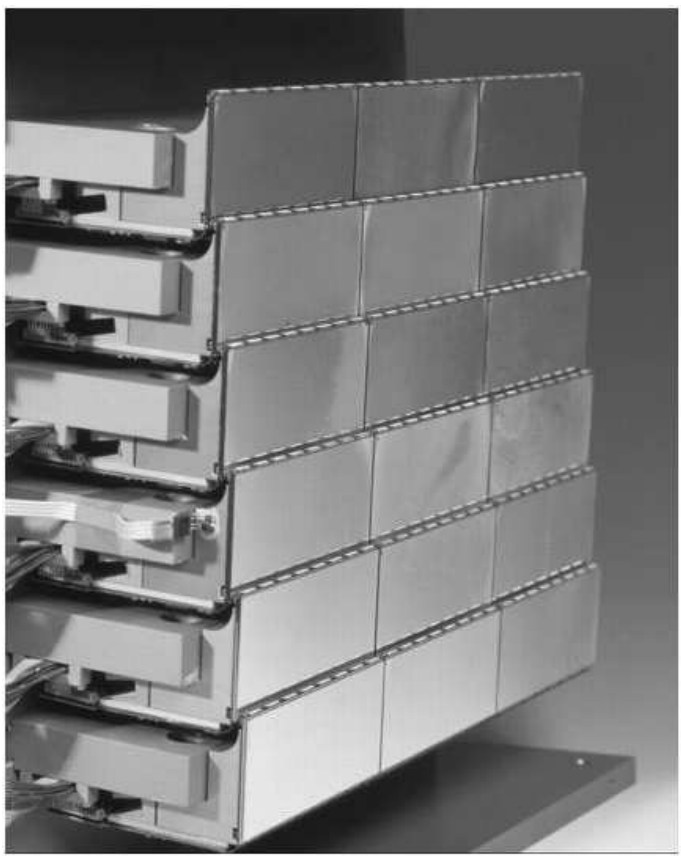}
\end{center}
\caption[]{Photograph of the 20x24 cm$^2$ large PILATUS 1M
detector for protein crystallography using counting hybrid pixel
detector modules.}
\label{pilatus}
\end{figure}

\begin{figure}
\begin{center}
\includegraphics[width=1.0\textwidth]{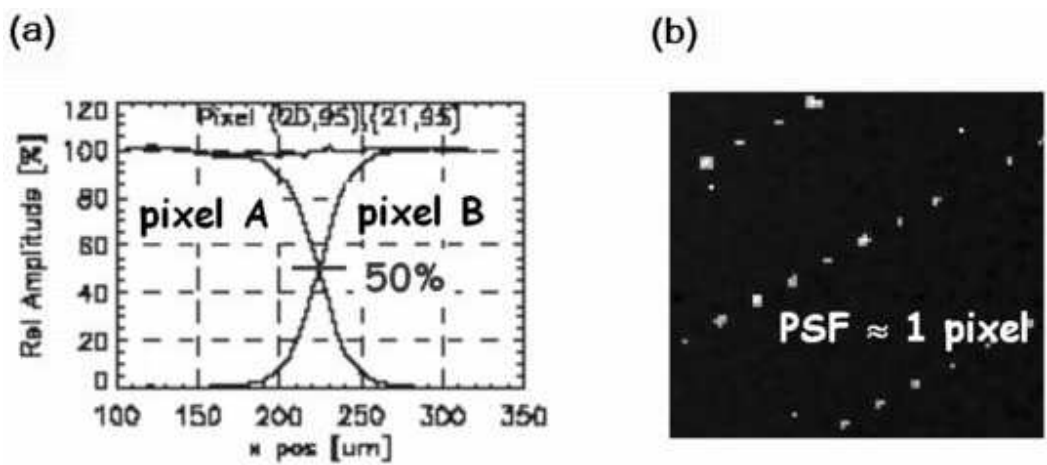}
\end{center}
\caption[]{(a) delicate threshold tuning with counting
pixel detectors at the borders in between pixels, (b) Bragg spots
of an image of Lysozyme taken with PILATUS 1M
\cite{PILATUS_Portland} are often contained in one pixel.}
\label{crystallography}
\end{figure}

Protein Chrystallography is another appealing application for 
counting pixel detectors. For the imaging of Bragg spots from
X-ray photons of $\sim$12 keV (corresponding to resolutions at 
the 1$\AA$ range) or higher, scattered off protein crystals, 
with high rate ($\sim$1-1.5 MHz/pixel) and high dynamic range 
\cite{Graafsma_Portland}, photon counting detectors have fundamental 
advantages over conventional integrating detector systems. That is
the high linearity of the counting method and the absence of so-called 
"blooming-effects", i.e. the response of non-hit pixels in the close 
neighborhood of a Bragg spot. For a typical Bragg spot the size
of a diffraction maximum is $100-200 \mu$m, calling for pixel sizes in
the order of $100-300 \mu$m, which is well achievable with today's
hybrid pixel detectors. A systematic limitation and
difficulty is the problem that homogeneous hit/count responses in
all pixels, also for hits at the pixel boundaries or between
pixels where charge sharing plays a role must be maintained by
delicate threshold tuning (Fig. \ref{crystallography}(a)).
Counting pixel developments are made for ESRF (Grenoble, France)
\cite{XPAD1,XPAD2} and SLS (Swiss Light Source at the
Paul-Scherrer Institute, Switzerland) beam lines. A photograph of
the PILATUS 1M detector \cite{PILATUS} at the SLS ($\sim$ 10$^6$
$217 \mu$m $\times 217 \mu$m pixels, 18 modules, 20$\times$24
cm$^2$ area) is shown in Fig. \ref{pilatus}. It is the
first large scale hybrid pixel detector in operation. Fig.
\ref{crystallography}(b) shows some Bragg spots obtained from a
Lysozyme crystal with 10s exposure to 12 keV sychrotron X-rays
\cite{PILATUS_Portland}.

\section{Trends in Hybrid Pixel Detectors}
With their mature technology Hybrid Pixel detectors prove to be the
State of the Art in 2D sensor systems. However there are some general
limitations and challenges for special applications. Hybrid Pixel
detectors are an expensive and complex technology, last but not least
due to yield issues of the many production steps (chip, sensor,
bumping, flip-chip). Also the achievable spatial resolution is limited to about $10 \mu$m
with pixel geometries of $100-300 \mu$m (with analog R/O). Where 
large area detectors are built, complex mechanical structures are
required to provide seamless coverage of the solid angle.
For imaging applications it is even more crucuial to use seamless
large area detectors without tiling to optimize the image reconstruction.
In addition, tracking applications impose a tight specification on
the material budget which is not optimal with Hybrid Pixel detectors.
Also the inevitable cooling for operation of Si detectors under
strong irradiation adds to that material budget.
Several ideas and developments are being pursued to address  
some of the issues stated above:

\subsection*{HAPS} Hybrid (Active) Pixel Sensors (HAPS)
\cite{HAPS} exploit capacitive coupling between pixels -- similar
to the same technique often used with silicon micro strip
detectors -- to obtain smaller pixel cells and pixel pitch with a
larger readout pitch resulting in interleaved pixels. The pixel
pitch is designed for best spatial resolution using charge sharing
between neighbors while the readout pitch is tailored to the needs
for the size of the front-end electronics cell. Sensor studies lead
to resolutions between 3$\mu$m and 10$\mu$m which could be obtained with
pixel (readout) pitches of 100$\mu$m (200$\mu$m). 

\subsection*{MCM-D}
\begin{figure}[htb]
\includegraphics[width=0.5\textwidth]{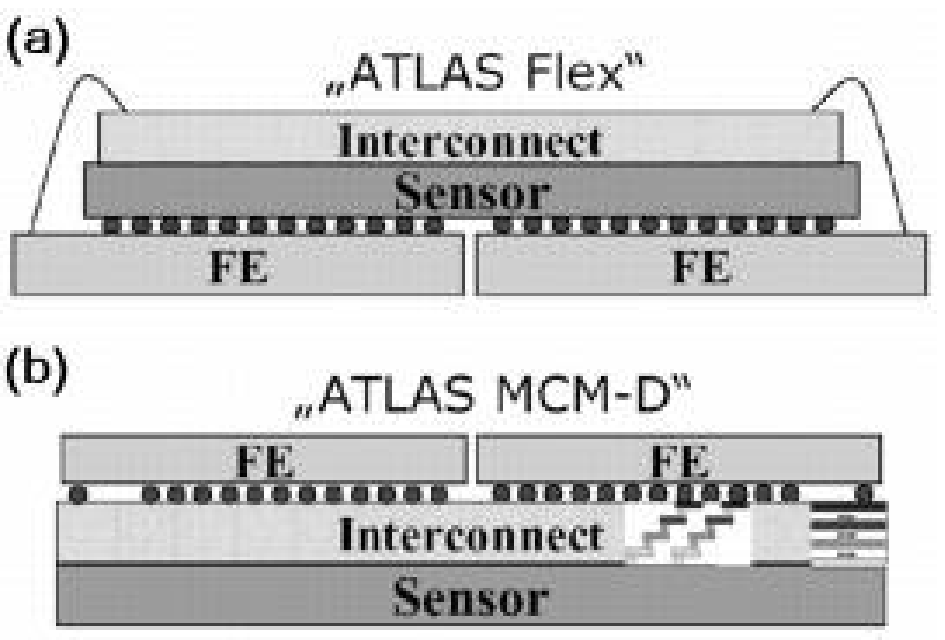}
\hskip 1cm
\includegraphics[width=0.3\textwidth]{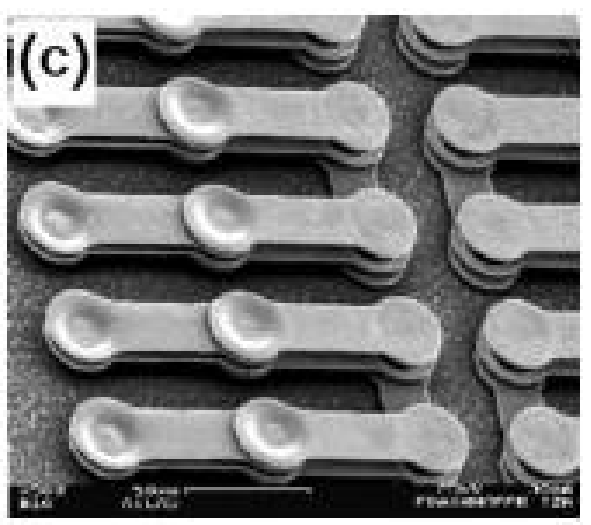}
\vskip 0.5cm
\begin{center}
\includegraphics[width=0.9\textwidth]{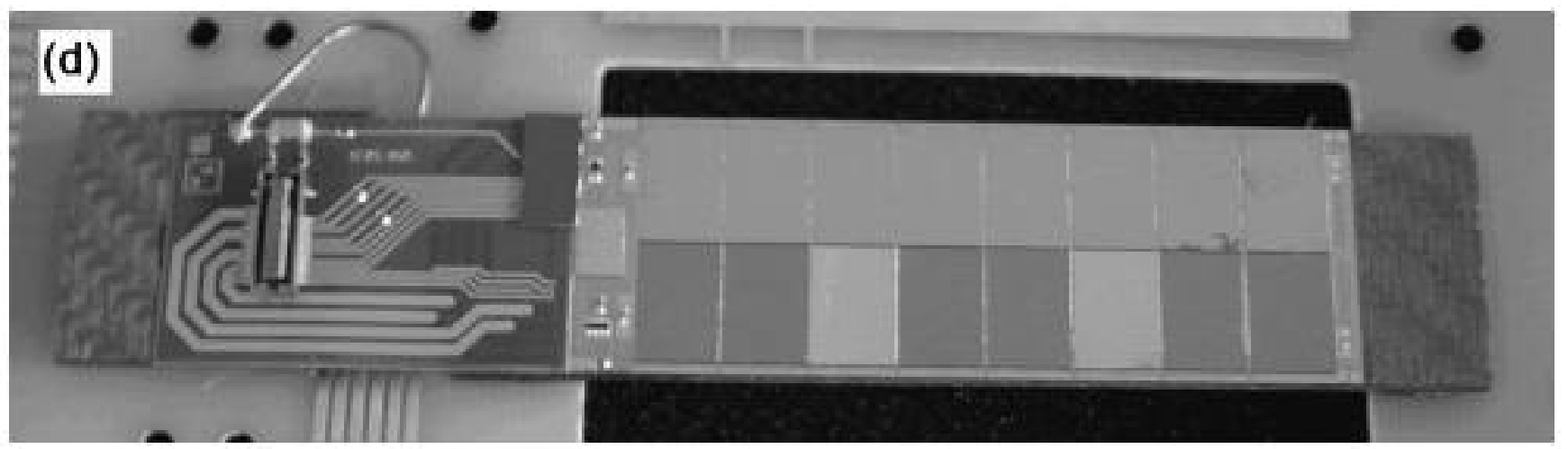}
  \caption{(top left, (a)) Schematic view of a hybrid pixel module and
  (top left, (b)) schematic layout of a MCM-D pixel module indicating the buried via
  structure, (top right) SEM photograph of a MCM-D via structure,
  (bottom) photograph of an ATLAS MCM-D module.
  }
  \label{MCMD}
\end{center}
\end{figure}

\begin{figure}
\begin{center}
\includegraphics[width=1\textwidth]{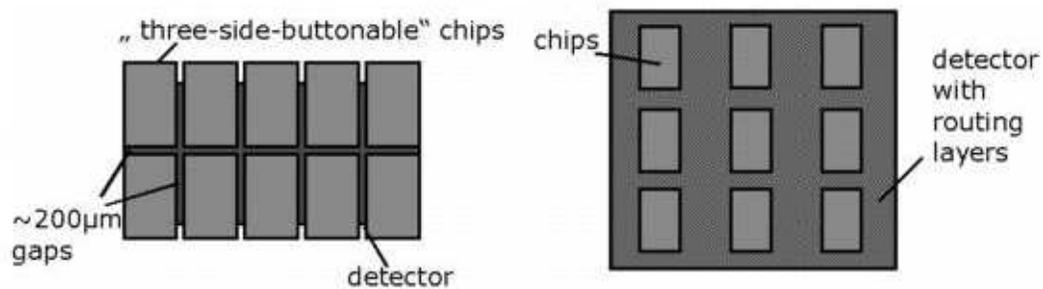}
\end{center}
\caption[]{(left) Schematic view of a conventional detector module build with 
"three-side-buttonable" chips. Gaps between chips are covered with larger sensor cells 
in those regions. The module geometry is limited to 2$\times$m chips per module. 
(right) Principle of an assembly with fan-out between electronic chips and sensor 
rendering seamless n$\times$m modules possible.}
\label{sparseCMOS}
\end{figure}

The present hybrid-pixel modules of the LHC experiments use an
additional flex-kapton fine-print layer on top of the Si-sensor
(Fig. \ref{MCMD}(a)) to provide power and signal distribution to
and from the module front-end chips. An alternative to the
flex-kapton solution is the so-called Multi-Chip-Module Technology
deposited on Si-substrate (MCM-D) \cite{MCMD1}. A
multi-conductor-layer structure is built up on the silicon sensor.
This allows to bury all bus structures in four layers in the
inactive area of the module thus avoiding the kapton flex layer
and any wire bonding at the expense of a small thickness increase
of $0.1 \% \ X_0$ (Fig. \ref{MCMD}(b)). The extra freedom in
routing also allows to design pixel detectors which have the same
pixel dimensions throughout the sensor. Fig. \ref{MCMD}(c) shows
a scanning electron microphotograph (courtesy IZM,
Berlin) of an MCM-D via structure, and Fig. \ref{MCMD}(d) shows
the photograph of an assembled ATLAS MCM-D module \cite{MCMD3}.
This concept also provides the possibility to build detector modules
which have non equal area ratios of sensor and electronic chips
when the routing layer implements a fan-out between the electronic pixels
and the corresponding sensor elements. With this approach it would 
be possible to build large area modules with n$\times$m chips and
almost 100\% active area (Fig. \ref{sparseCMOS}).

\subsection*{3-D Silicon Detectors, Active Edge}
\begin{figure}
\begin{center}
\includegraphics[width=0.5\textwidth]{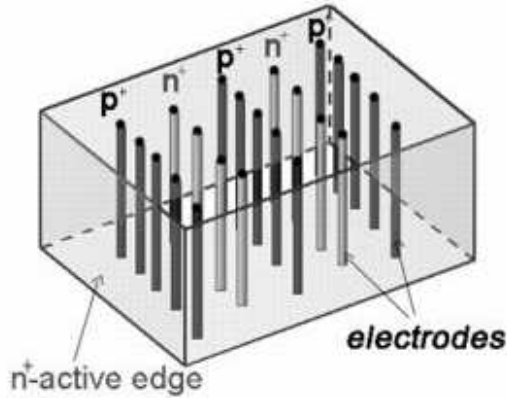}
\end{center}
\caption[]{Structure of a 3-D detector. The needle shaped p+ and n+ electrodes 
are fabricated from etched holes which are subsequently filled with appropriate 
doped polysilicon. Their typical pitch is 50 $\mu$m.}
\label{3d}
\end{figure}

So called 3-D detectors provide a lateral drift field between their needle like 
alternating p+ and n+ electrodes which have a typical pitch of 50$\mu$m \cite{3D-parker} 
(Fig. \ref{3d}). Thus they exhibit a very fast charge collection (1 - 2 ns) at low 
depletion voltages ($<$ 10 V) which make them well suited for harsh radiation environments. 
They are build using micromechanical systems technology involving
the use of support wafers and reactive ion etching which makes their fabrication
more complex compared to standard planar processes. The same technology can be used
to extend the sensitive area of a detector within 10~$\mu$m of its edge which
leads to a so called active-edge detector \cite{active_edge}.

\section{Monolithic and Semi-Monolithic Pixel Detectors}
The ultimate goal of a detector development would be the 
integration of sensor, amplifying electronics and read-out 
logic on one single substrate in a technology which would
be commercially available. So far different attempts to develop 
such monolithic pixel detectors have been undertaken. Much of
these efforts are influenced by R$\&$D for vertex tracking
detectors at future colliders such as a International Linear $e^+e^-$
Collider (ILC) \cite{TESLA-TDR}. These detectors require a
very low material budget per layer ($\ll$1$\%$ X$_0$), small 
pixel sizes ($\sim$20$\mu$m$\times 20 \mu$m) and a very high rate capability
(80 hits/mm$^2$/ms) yielding challenging requirements on the
sensor and the electronics circuitry.

The different monolithic approaches can be characterized according 
to some fundamental features regarding the charge collection process
and the complexity of electronic circuitry. One distinction is based upon the fact
whether full CMOS circuitry is allowed (also in the pixel active area)
and if the approach is based on or at least compatible with a standard 
process technology. The other basic criterion is the charge collection 
process and thus the amount (and uniformity) of the generated signal.
Charge collection occurs either in a fully depleted bulk providing a large signal or 
in a thin undepleted epi-layer, respectively amorphous-Si layer, which leads to smaller signals.
The above mentioned ultimate monolithic goal would be fulfilled
with a full CMOS commercial standard device with charge collection
in a fully depleted bulk.

\begin{itemize}
\item \emph{\bf Non-standard CMOS on high resistivity bulk} \\
The first monolithic pixel detector was successfully operated in a
particle beam already in 1992 \cite{parker92} using a high
resistivity p-type bulk p-i-n detector in which the junction had
been created by an n-type diffusion layer. On one side, an array
of ohmic contacts to the substrate served as collection
electrodes. Due to this only pMOS transistor circuits sitting in
n-wells were possible to be integrated in the active area. The
technology was certainly non-standard and non-commercial. No
further development emerged.
\hfill \\

\item \emph{\bf Standard CMOS technology with charge collection in epi-layer} \\
Certain CMOS technologies offer a lightly doped epitaxial silicon layer
of a few to 15$\mu$m thickness between the low resistivity silicon
bulk and the planar processing layer which can be used for charge
collection \cite{meynants98,MAPS1,MAPS2}. The generated charge is
kept in a thin epi-layer atop the low resistivity silicon bulk by
potential wells at the boundary and reaches an n-well collection
diode by thermal diffusion (cf. Fig. \ref{MAPS}(a)). The sensor is
depleted only directly under the n-well diode. The signal charge
is hence very small ($<$1000e) and mostly incomplete; low noise
electronics is the challenge in this development. Due to the diffusion
process the charge collecting time might also be an issue for
high rate applications.
Collaborating groups around IReS$\&$LEPSI \cite{LEPSI1,LEPSI-Portland}, RAL
\cite{RAL-Vertex03} and Irvine-LBNL-Ohio \cite{kleinfelder03} use
similar approaches to develop large scale CMOS active pixels also
called MAPS (Monolithic Active Pixel Sensors) \cite{MAPS1}.
Prototype detectors have been produced in $0.6 \mu$m, $0.35 \mu$m
and $0.25 \mu$m CMOS technologies \cite{Dulinski03,AGay03}.

Matrix readout of MAPS is performed using a standard 3-transistor
circuit (line select, source-follower stage, reset) commonly
employed by CMOS matrix devices, but can also include current
amplification and current memory \cite{Dulinski03}. For an image
two complete frames are subtracted from each other (correlated double 
sampling, CDS) which suppresses switching and low frequency noise. 
Noise figures of 10-30e and S/N $\sim$20 have been achieved with 
spatial resolutions below 5$\mu$m. Regarding radiation hardness MAPS 
appear to sustain non-ionizing radiation (NIEL) to $\sim$10$^{12}$n$_{eq}$ while the
effects of ionizing radiation damage (IEL) are at present still
under investigation. The present focus of further development lies
in making larger area devices for instance by stitching over
reticle boundaries \cite{AGay03}, increasing the charge collection
performance in the epi-layer by triple-well \cite{RAL-Vertex03},
photo-gate \cite{Kleinfelder-Portland}, and photo-FET
\cite{Dulinski03,LEPSI-Portland} techniques and developing a
higher radiation tolerance. In addition, for applications like
e.g. precise beam position monitoring in hadron therapy, devices
with very thin entrance windows are needed to detect $\sim$20 keV
electrons scattering off a thin metal foil held in the hadron
beam. Such a thinned MAPS detector, which is also capable of
autoradiographic tritium detection, is shown in Fig.
\ref{MAPS}(b).

\begin{figure}[h]
\begin{center}
\includegraphics[width=0.6\textwidth]{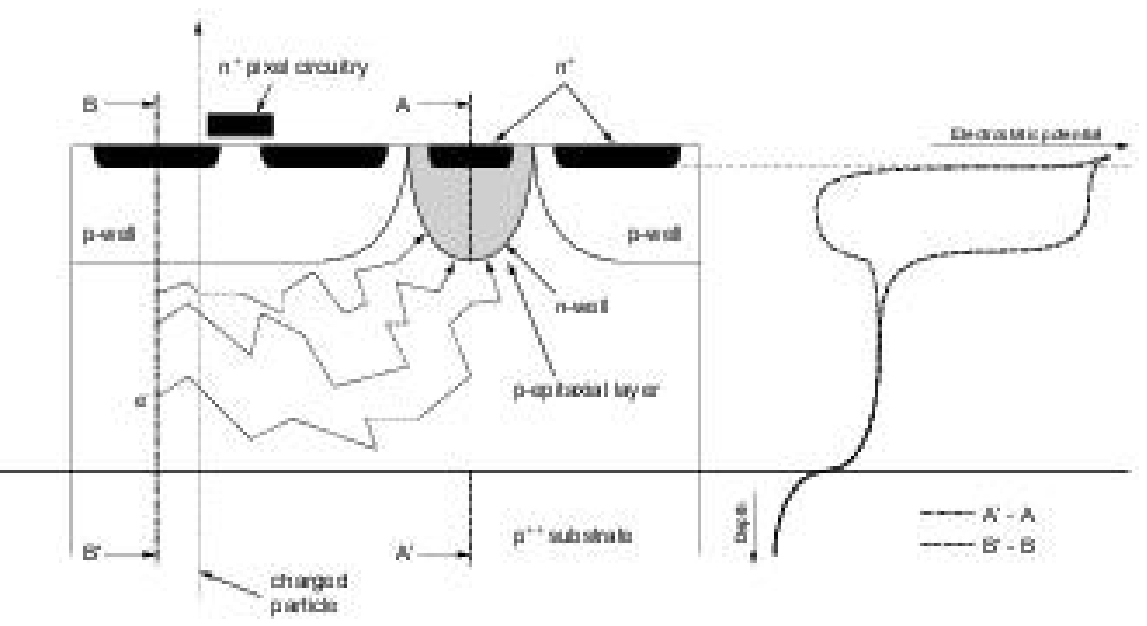}
\hskip 0.5cm
\includegraphics[width=0.35\textwidth]{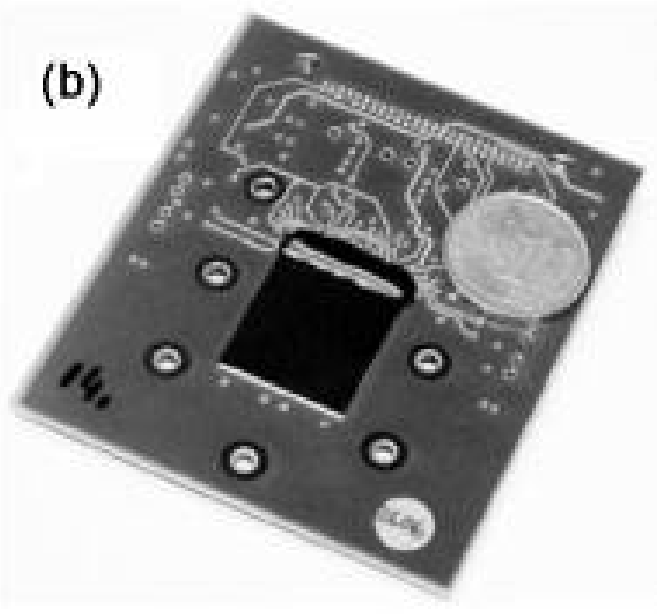}
\end{center}
\caption[]{(a) Principle of an Monolithic Active Pixel Sensor
(MAPS) targeting CMOS electronics with low
resistivity bulk material. The charge is generated and collected
by diffusion in the few $\mu$m thin epitaxial Si-layer. (b) MAPS
detector with 100 nm thin backside entrance window.} \label{MAPS}
\end{figure}

Being a standard CMOS process this approach has the potential to build 
large area detectors at low costs ($\sim$25$\$$ per cm$^2$). On the other 
hand the thickness of the epi-layer is technology dependent and scales with the decreasing 
feature size of future processes hence further reducing the available signal charge. 
Another drawback of this approach is the fact that despite using CMOS technology the 
potential of full CMOS circuitry in the active area is not available (only nMOS) because 
of the n-well/p-epi collecting diode which does not permit other n-wells. 
Nevertheless improved readout concepts and device development for high rate particle 
detection at a linear collider are under development \cite{TESLA-TDR}. As a concrete 
project for an experiment the STAR micro vertex detector upgrade plans the use
of CMOS active pixels \cite{Dulinski03}.

\hfill \\
\item \emph{\bf Non-standard SOI on high resitivity bulk} \\
To join the features of full charge collection and the utilization of 
true full CMOS circuitry the authors of \cite{SOI-Portland} propose
the development of sensors using a silicon-on-insulator (SOI) wafer
with an high resestivity bulk material. SOI process technology which has
been developed in favor for advanced high-speed and low-power 
circuits uses special wafers with a thin monolithic Si layer (50 nm -- 1.5 $\mu$m)
atop a buried oxide (SiO$_{2}$) layer. This layer effectively isolates
the electronically active SOI layer from the bulk material rendering it a pure 
mechanical support. The idea of this development approach is to
use a high resistivity material for the support wafer to achieve 
a fully depletable charge generation layer of suitable thickness (200--300$\mu$m).
The charge collection will be obtained by vias which connect the 
high resitivity bulk with electronic circuits through the insulation layer (Fig. \ref{SOI}). 
Present results are based also on a non-standard SOI technology 
but this interesting development is still in its beginnings and might have the 
potential to become compatible with commercial process technologies \cite{SOI_Hiroshima}.

\begin{figure}[h]
\begin{center}
\includegraphics[width=0.75\textwidth]{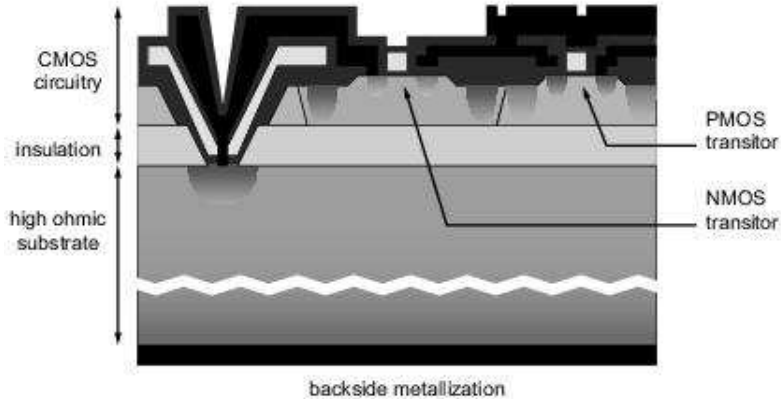}
\end{center}
\caption[]{Cross section through a monolithic CMOS on SOI
pixel detector using high resistivity silicon bulk insulated from
the low resistivity CMOS layer with connecting vias in between
\cite{HAPS-SOI,SOI-Portland}.} \label{SOI}
\end{figure}

\hfill \\
\item \emph{\bf Amorphous silicon on standard CMOS ASICs} \\
A development approach which is compatible with standard CMOS
technology is based on the deposition of the sensor material
as an amorphous film direct on the surface of the readout ASIC.
Hydrogenated amorphous-Silicon (a-Si:H) has been studied as a 
sensor material long ago and has gained interest again \cite{theil2001,jarron02} 
with the advancement in low noise, low power electronics.
With the deposited film being only $<$30$\mu$m thick the collected
charge is in the range of only 500-1500 electrons which 
makes ultra low-noise electronic inevitable. A cross section through a 
typical a-Si:H device is shown in Fig. \ref{aSi}. Although not a
entirely monolithic concept the amorphous silicon on standard CMOS ASICs
approach lacks the demanding process of hybridization. Also
the radiation hardness of these detectors appears to be very high $>$10$^{15}$cm$^{-2}$ 
due to the defect tolerance and defect reversing 
ability of the amorphous structure and the larger band gap (1.8 eV). For
high-Z applications poly-crystalline HgI$_2$ constitutes a
possible semiconductor film material. The potential advantages are
small thickness, radiation hardness, and low cost. The development
is still in its beginnings and the main challenge -- beneath low noise
VLSI design -- is the quality of the amorphous material and the technology
of its deposition.

\begin{figure}[h]
\begin{center}
\includegraphics[width=0.65\textwidth]{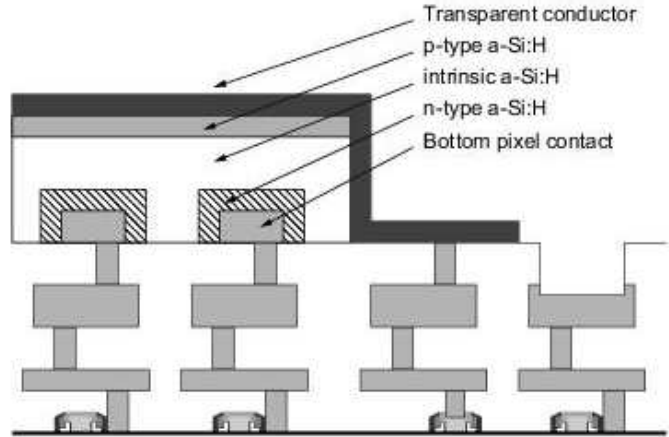}
\end{center}
\caption[]{Cross section through a structure using amorphous silicon on top of standard 
CMOS VLSI electronics \cite{theil2001,jarron02}.} \label{aSi}
\end{figure}

\hfill \\
\item \emph{\bf Amplification transistor implanted in high resistivity bulk} \\
The basic idea of this development approach is the integration of 
the first amplifying transistor of the readout electronic into the sensor
substrate. This so-called DEPFET pixel sensors \cite{kemmer87} have 
a JFET or MOSFET transistor implanted in every pixel on a sidewards depleted
\cite{gatti84} bulk. Electrons generated by radiation in the bulk
are collected in a potential minimum underneath the transistor. This 
potential minimum acts as an internal gate and the collected charge is 
modulating the channel current (Fig. \ref{DEPFET_principle}).
With additional contacts of the external gate and a so-called clear contact
individual transistors in a 2-D matrix can be selected for readout and
subsequently reseted. The bulk is fully depleted rendering large signals and 
at the same time the small capacitance of the internal gate offers low noise operation. 
With round single pixel structures noise figures of 2.2e at room temperature
and energy resolutions of 131 eV for 6 keV X-rays have been obtained
\cite{depfet_IEEE03}. This features make DEPFET pixel detectors a competitive 
sensor system for applications which demand excellent energy resolution like X-ray 
astronomy \cite{holl02} or biomedical autoradiography \cite{ulrici03}.
Also for the use as a vertex detector at the Future Linear Collider the DEPFET 
pixel detectors are under investigation \cite{DEPFET-TESLA,depfet_IEEE03}. 
They would allow for thinned detectors ($\sim$50$\mu$m \cite{laci04}) to comply with 
the tight material budget and a very low power operation of a row-wise switched matrix 
\cite{Trimpl02,depfet_IEEE03}. The sensor technology is non-standard and the operation of 
DEPFET pixel detectors requires separate steering and amplification ICs.

\begin{figure}[h]
\begin{center}
\includegraphics[width=0.8\textwidth]{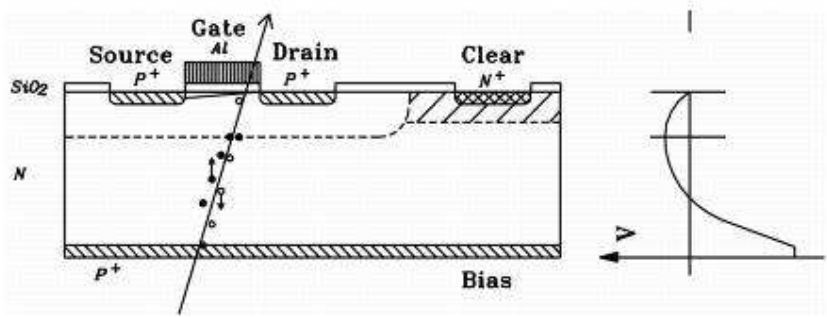}
\end{center}
\caption[]{Principle of operation of a DEPFET pixel structure
based on a sidewards depleted detector substrate material with an
embedded planar field effect transistor. Cross section (left) of
half a pixel with symmetry axis at the left side, and potential
profile (right).} \label{DEPFET_principle}
\end{figure}

\begin{figure}[htb]
\begin{center}
\includegraphics[width=1.0\textwidth]{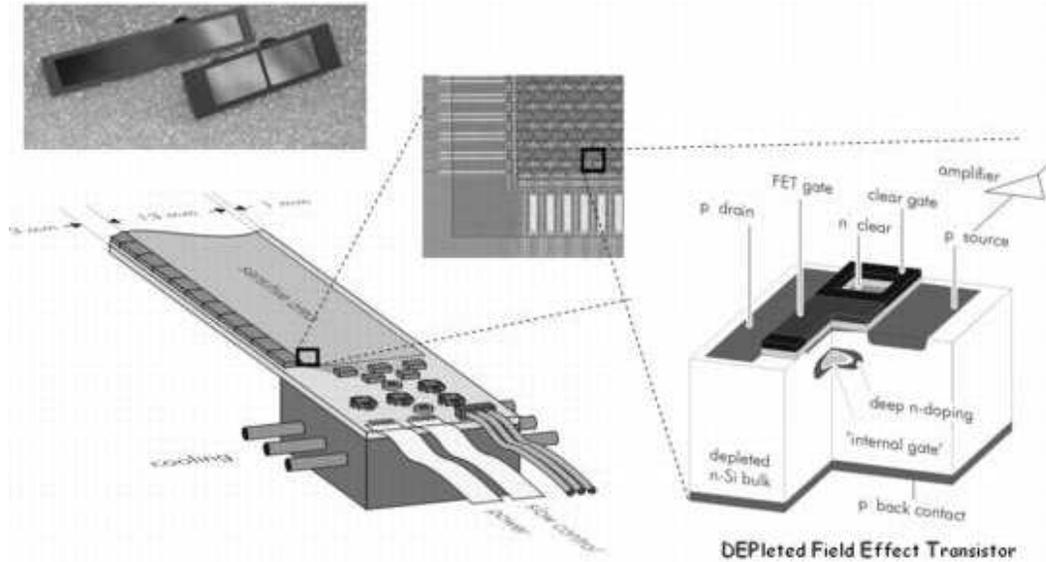}
\end{center}
\caption[]{Sketch of a TESLA first layer module with thinned
sensitive area supported by a silicon frame. The enlarged view
show a DEPFET matrix and a DEPFET double pixel structure (20x30$\mu$m$^2$ pixel size),
respectively. The photo (upper left) shows silicon diode
structures thinned to 50$\mu$m thickness by anisotropic etching
\cite{laci04} in a 300$\mu$m thick frame.} \label{DEPFET_TESLA}
\end{figure}

\end{itemize}

\section{Summary}
The development of Hybrid Pixel technology had its origin in the demand 
for high performance vertex detectors for particle physics experiments.
Along with it came large advancements in the development of low noise electronics,
hybridization technology and understanding and control of radiation 
effects in sensor material and readout electronics. Also new
application areas like X-ray astronomy, medical imaging and
protein chrystallography emerged which require well adapted electronic 
concepts to cope with the varying demands on energy resolution and rate 
capability. This new range of applications and the increasing demand on the  
performance of future vertex detectors leads to new development trends. Some of them 
are enhancements to the Hybrid Pixel technology, as the sensor and electronic chip
are still different entities: MCM-D structures allow large area modules
and 3-D sensors with active edges improve the active/inactive area ratio and
the radiation tolerance. Other interesting developments lead to monolithic
detetion systems: MAPS sensors which employ the epi-layer of a
standard CMOS technology for signal detection, the deposition 
of thin films like amorphous-Si on CMOS ASICs and the use of a
CMOS-SOI process on a wafer with high resistivity bulk material.
A semi-monolithic approach are the DEPFET pixels which integrate active 
transistors on a high resistivity bulk which are steered via
external ASICs.

\bibliographystyle{unsrt}
\bibliography{EMRS04}

\end{document}